\documentclass{aa}

\input epsf 

\begin{document}

\sloppy

\thesaurus{3 (11.02.2, 13.07.2)} 

\title{Search for a TeV gamma-ray halo of Mkn 501}

\authorrunning{F. Aharonian et al.}

\author{F.A. Aharonian\inst{1},
A.G.~Akhperjanian\inst{7},
J.A.~Barrio\inst{2,3},
K.~Bernl\"ohr\inst{1},
O.~Bolz\inst{1},
H.~B\"orst\inst{5},
H.~Bojahr\inst{6},
J.L.~Contreras\inst{3},
J.~Cortina\inst{2},
S.~Denninghoff\inst{2}
V.~Fonseca\inst{3},
J.C.~Gonzalez\inst{3},
N.~G\"otting\inst{4},
G.~Heinzelmann\inst{4},
G.~Hermann\inst{1},
A.~Heusler\inst{1},
W.~Hofmann\inst{1},
D.~Horns\inst{4},
A.~Ibarra\inst{3},
C.~Iserlohe\inst{6},
I.~Jung\inst{1},
R.~Kankanyan\inst{1,7},
M.~Kestel\inst{2},
J.~Kettler\inst{1},
A.~Kohnle\inst{1},
A.~Konopelko\inst{1,}$^\S$,
H.~Kornmeyer\inst{2},
D.~Kranich\inst{2},
H.~Krawczynski\inst{1},
H.~Lampeitl\inst{1},
E.~Lorenz\inst{2},
F.~Lucarelli\inst{3},
N.~Magnussen\inst{6},
O.~Mang\inst{5},
H.~Meyer\inst{6},
R.~Mirzoyan\inst{2},
A.~Moralejo\inst{3},
L.~Padilla\inst{3},
M.~Panter\inst{1},
R.~Plaga\inst{2},
A.~Plyasheshnikov\inst{1,}$^\S$,
J.~Prahl\inst{4},
G.~P\"uhlhofer\inst{1},
W.~Rhode\inst{6},
A.~R\"ohring\inst{4},
G.P.~Rowell\inst{1},
V.~Sahakian\inst{7},
M.~Samorski\inst{5},
M.~Schilling\inst{5},
F.~Schr\"oder\inst{6},
M.~Siems\inst{5},
W.~Stamm\inst{5},
M.~Tluczykont\inst{4},
H.J.~V\"olk\inst{1},
C.~Wiedner\inst{1},
W.~Wittek\inst{2}}

\institute{Max Planck Institut f\"ur Kernphysik,
Postfach 103980, D-69029 Heidelberg, Germany \and
Max Planck Institut f\"ur Physik, F\"ohringer Ring
6, D-80805 M\"unchen, Germany \and
Universidad Complutense, Facultad de Ciencias
F\'{i}sicas, Ciudad Universitaria, E-28040 Madrid, Spain 
\and
Universit\"at Hamburg, II. Institut f\"ur
Experimentalphysik, Luruper Chaussee 149,
D-22761 Hamburg, Germany \and
Universit\"at Kiel, Institut f\"ur Experimentelle und Angewandte Physik,
Leibnizstra{\ss}e 15-19, D-24118 Kiel, Germany\and
Universit\"at Wuppertal, Fachbereich Physik,
Gau{\ss}str.20, D-42097 Wuppertal, Germany \and
Yerevan Physics Institute, Alikhanian Br. 2, 375036 Yerevan, 
Armenia\\
\hspace*{-4.04mm} $^\S\,$ On leave from  
Altai State University, Dimitrov Street 66, 656099 Barnaul, Russia\\
}

\mail{Werner Hofmann, \\Tel.: (Germany) +6221 516 330,\\
email address: Werner.Hofmann@mpi-hd.mpg.de}

\offprints{Werner Hofmann}

\date{Received ; accepted }

\maketitle

\begin{abstract}

For distant extragalactic sources of gamma-rays in the PeV 
($10^{15}$ eV) energy
range, interactions of the gamma rays with intergalactic diffuse
radiation fields will initiate a pair cascade. Depending on the magnetic
fields in the vicinity of the source, 
the cascade can either result in an isotropic halo around 
an initially beamed source, or remain more or less collimated.
Data recorded by the HEGRA system of imaging atmospheric Cherenkov
telescopes are used to derive limits on the halo flux from the AGN 
Mrk 501. This is achieved by
comparing the angular distribution of TeV gamma-rays 
during the 1997 burst phase -- where direct photons should
dominate -- with the distribution during the 1998/99 quiescent 
state, where a steady-state halo contribution should be most pronounced.
The results
depend on the assumptions concerning the angular distribution of the
halo; limits on the halo flux within $0.5^\circ$ to $1^\circ$ from
the source range between 0.1\% and 1\% of the peak burst flux.

\keywords{galaxies: BL Lacertae objects: individual: Mkn 501 -
gamma rays: observations}

\end{abstract}

\section{Introduction}

High-energy gamma rays have a limited mean free path in the universe,
due to interactions with the intergalactic soft photon radiation,
resulting in pair production
(Nikishov 1962,
Goldreich \& Morrison 1964,
Gould \& Schreder 1966, Jelley 1966, Stecker et al. 1992). 
For (head-on) collisions of a gamma
ray of energy $E$ with a photon of wavelength $\lambda$,
the pair production threshold is crossed if 
$\lambda < 2.4~E \mbox{($\mu$m/TeV)}$. At PeV energies, interactions with the
microwave background radiation limit the mean free path of 
gamma rays to galactic
distance scales.
At lower (TeV) energies, infrared or optical 
photons are required to cross the pair production threshold, and 
the mean free path rapidly grows 
with decreasing energy to Mpc scales and beyond. 
When a source injects very high-energy gamma-rays into this 
photon background, a cascading process will start, where 
the gamma ray converts into a pair, the electron and positron
in turn Compton-scatter background photons to high energy, 
which again produce a pair, and so on until the energies are 
so low that the medium
is transparent to these gamma-rays, on the scale of
the distance between source and observer
(Aharonian \& Atoyan 1985, 
Protheroe 1986, Zdziarski 1988, Protheroe \& Stanev 1993,
Blandford \& Levinson 1995). The cascading process
results in universal photon energy spectra, almost
independent of the energy of the primary photons. The cascade photons
modify the spectrum of direct photons from the source, since 
cascade photons pile up around the energy where the medium
becomes transparent. In essence,
the energy content in very high-energy photons is transposed
into photons of lower energy. To which extent the spectrum
of direct photons is modified by the cascade photons depends
on the spectral index and on the high-energy cutoff of the primary radiation.
If the source emits well into the PeV region, and if the
spectrum resembles a power law $\mbox{d}N/\mbox{d}E \sim E^{-\alpha}$
with index $\alpha \approx 2$ -- with equal
amount of power per decade of energy -- the yield of cascade 
photons is comparable to the number of direct photons. If, 
on the other hand, the source cuts off at lower energies, or
if the primary spectrum is steeper, the energy content of the high-energy
photons is small compared to the energy density at lower energies,
and the cascade photons represent only a small correction
to the spectrum. Already an index of 2.2 suffices to reduce the
influence of cascade photons dramatically
(Protheroe \& Stanev 1993). 

While the discussion
given so far concerns the {\em spectral distribution}
of all photons produced by the source and the cascading effect, 
relevant e.g. for the description of the diffuse 
gamma-ray background resulting from the superposition of many unresolved point
sources, cascading is also relevant for 
the {\em angular distribution} of photons from individual point sources
and may give rise to a halo surrounding distant high-energy sources
such as AGNs (Aharonian et al. 1994, 1995). 
If the magnetic fields in the vicinity 
of the source are such that the average mean free path between
cascade steps -- at least during the initial stages of the 
cascade -- is large compared to Larmor radii, the cascade 
will generate, even for highly beamed primary sources, an isotropic
halo. 
Concerning the intergalactic part of the pair cascade,
the required fields are in the range of some nG to some ten nG
(Aharonian et al. 1994),
and while little is known about intergalactic magnetic fields,
such values are not implausible, and are seen at least in the 
halos of galaxies and in clusters of galaxies (Kronberg 1996,
Eilek 1999).
For sources such as the AGNs Mkn 501 or 421, with 
characteristic distances around 140 Mpc
(for $H$ = 65 km/s/Mpc), halo sizes have been estimated
to a few degrees (Aharonian et al. 1994, 1995). 

Since the isotropic cascade 
proceeds over large, typically Mpc scales,
the halo intensity will be constant in time,
compared to the variable nature of the direct radiation. 
 From simple geometrical arguments, one finds that the characteristic
time scale $T$ governing the evolution of the 
halo flux seen at an angular distance
$\theta$ from a source at a distance $d$ is given by
$$
T \approx {\theta d \over c} \approx 4 \cdot 10^5~\mbox{y}
$$
in case the halo is generated in the vicinity of the source;
for the numerical estimate $\theta = 0.05^\circ$ is used, the
smallest scale where a halo might be detectable, given the angular
resolution of the instrument, and $d$~=~140~Mpc. The
shortest time scale is possible if the cascades are generated 
half-way between the source and the observer, in a region of small
magnetic fields and low deflection angles,
$$
T \approx {\theta^2 d \over 2 c} \approx 150~\mbox{y}~~~.
$$
One
might, therefore, 
spec\-ulate that the very low quiescent-phase TeV gamma-ray 
flux from AGNs such as Mkn 421 or 501 represents the halo flux,
onto which direct photons from burst events are superimposed. In
such a scheme, the ratio of the isotropic halo flux $F_{Halo}$ 
to the typical flux in bursts $F_{Burst}$ can be estimated as
$$
F_{Halo} / F_{Burst} \approx r~D~f
$$
where $r$ is the ratio of the energy flux in the PeV and TeV energy ranges 
provided by the source, $D$ is the duty cycle, i.e. the fraction of the
time the source is in the `burst' state, and $f$ is the beaming fraction, 
assuming that the
observer is located in the center of the beam for burst events.
Assuming
 a flat spectral energy distribution of the source
($r \approx 1$), a beaming fraction $O(1/100)$ and a duty cycle
of $O(1/10)$ -- one predicts a halo flux which is at least a factor of
1000 below the peak flux. On the other hand, if recent data
on the strength of the infrared background and their implication
concerning the source spectra are taken literally -- see e.g. 
Protheroe \& Meyer (2000) --
reconstructed source spectral energy distributions rise steeply
at high energy, implying very large values of $r$, $O(10^2)$ to 
$O(10^4)$ and a halo flux comparable to the burst flux.
Also, smaller intergalactic magnetic fields 
could result in a beamed halo and a corresponding
increase in halo flux. In the extreme case, with fields as low as
$10^{-20}$ G, as discussed by Plaga (1995) and Kronberg (1995),
the intergalatic part of the cascade is so well collimated that for all
practical purposes it cannot be resolved.

This paper describes a search for a TeV gamma-ray halo surrounding
Mkn 501, using the large data base accumulated with the HEGRA
system of imaging atmospheric Cherenkov telescopes in the years
1997 through 1999.
While Mkn 501 is not the ideal candidate for a halo source --
given its relatively modest luminosity and the fact that the
`ideal' distance to observe halos is in the range of several
100 Mpc to 1 Gpc (Aharonian et al. 1994), the uncertainties
in the model expectation are significant and motivate an experimental search
for a gamma-ray halo. 

\section{The Mkn 501 data set and the event selection}

The AGN Mkn 501 has been observed on a regular basis during the
observation (visibility) periods since
1997 using the HEGRA system of five imaging atmospheric
Cherenkov telescopes, located on the Canarian Island of La Palma
on the site of the Observatorio del Roque de los Muchachos.
The HEGRA telescope system allows the stereoscopic 
reconstruction of air showers above a threshold of 500~GeV,
and provides event-by-event a determination of the photon
direction with a precision of typically $0.1^\circ$, and of
the photon energy with a resolution of better than 20\%
(Konopelko et al. 1999).

The following analysis is based on the 1997 
data set, acquired during the burst phase of Mkn 501 and 
described in detail
by Aharonian et al. (1999a, 1999b), and on the 1998/1999 
quiescent-phase data set, 
described by Aharonian et al. (2000a). With exceptions detailed
below, the same methods for data selection and event reconstruction
were applied to the data sets in this analysis.

In order to search for halo components, it is important to
control the point spread function of the instrument and to
reduce as much as possible tails in the 
point spread function. Therefore,
the techniques described by Hofmann et al. (2000)
were applied to reconstruct the
shower geometry from the observed Cherenkov images, providing,
for each shower, an error estimate for the direction of the
photon. Well-reconstructed photons were then selected by 
requiring an error of less than $0.1^\circ$ in 
each projection, keeping slightly more than
 50\% of all photons. The same approach,
with an even tighter selection, has already been applied in a study
of the size of the TeV gamma ray source in the Crab Nebula
(Aharonian et al., 2000b).

Past HEGRA studies of the TeV gamma radiation from Mkn 421 and Mkn 501
(Aharonian et al. 1999a, 1999b, 1999c, 2000a)
or the Crab nebula (Aharonian et al. 1999d)
used rather loose cuts to select gamma ray 
events, in order to minimize systematic errors. For the present study,
to gain maximal sensitivity for a faint halo, tighter cuts were
applied; the basic idea is to determine the ratio of halo 
intensity to direct flux with tight cuts, and then derive an
absolute  
halo flux by multiplying this ratio with the absolute flux determined
with loose cuts for the direct radiation. 
Gamma-ray selection cuts are based
on the {\it mean scaled width} of images 
(Aharonian et al. 1999a), where gamma-rays 
appear as a roughly Gaussian peak at 1 with a width of about 10\%,
whereas nucleonic cosmic rays generate a broad distribution
at larger {\it mean scaled width} values.
Both simulations and Crab and Mkn 501 data show that the best
sensitivity for faint sources is achieved with a cut requiring a 
{\it mean scaled width} below 1.05 to 1.10, compared to cuts at
1.2 to 1.3 in analyses aiming at a precision determination of
the flux and spectrum of strong sources. Here, a cut at 1.1
is used, significantly increasing the rejection of 
cosmic-ray induced events. The slightly increased systematic errors in the
determination of the absolute flux
\footnote{ For the selection of events used in this analysis,
the cut at 1.1 retains $(84 \pm 5) \%$ of all gamma rays, compared
to $(97 \pm 2) \%$ for a cut at 1.2, and 99\% at 1.3.} are still negligible in the
determination of flux limits, and in any case cancel to a large
extent when the ratio of the halo flux to the central direct flux is
taken.

\section{Search for a gamma-ray halo}

Depending on the angular distribution of the halo radiation, a 
halo can manifest itself in different ways in observations with
Cherenkov telescopes:
\begin{itemize}
\item A component of the halo whose angular spread is very small
compared to angular resolution of the instrument - about 4' for 
the data set discussed here - will be superimposed onto the
direct radiation and cannot be resolved. This limits in particular
the detection of the collimated halos generated by cascades
in regions of very small magnetic fields.
\item A component with an angular spread comparable to the
instrumental resolution will result in a widening of the
angular distribution, without explicitly resolving the halo.
\item A halo with an angular distribution which is wide compared
to the resolution can be detected directly; since much of the
directional signature is lost in this case, the evaluation of the
background due to cosmic rays becomes then
a critical issue.
\end{itemize}
The background mentioned in the third scenario results from
the small fraction of charged cosmic rays
with a shower evolution similar to that of gamma-ray showers. Since it
is uncertain if this remaining background can be modeled with 
sufficient precision by the simulations, frequently an off-source region
is used to measure the background. In case of the HEGRA telescopes,
the source is placed $0.5^\circ$ off axis in declination, and a region 
displaced by $0.5^\circ$ in the opposite direction is used as a 
background region. The sign of the displacement alternates
from run to run, to minimize systematic effects. Compared to
dedicated off-source runs, this method has the advantage that
the off-source data are taken at the same time and under 
exactly the same weather conditions and operating
conditions as the on-source data; in addition,
the on-source observation time is maximized. A disadvantage 
of the technique lies in the small separation of $1^\circ$ between
the on-source and off-source region. For a source extended 
on this scale, the off-source region also contains genuine
gamma rays, and the background subtraction reduces the signal.
A diffuse source generating a uniform distribution cannot
be detected by this method. For a halo with an extent of
a few degrees, this problem can be circumvented by using background
data obtained from a different region of the sky; since these data
are usually not recorded under identical conditions, systematic
uncertainties in the background estimate will increase.

In the following, we will first examine, with the 
standard background subtraction, the angular distribution of gamma rays,
searching for
qualitative indications of a halo. We then use models for the halo
angular distribution to quantify the results. Finally, the halo
flux is recalculated using a distant background region to cover 
the case of a halo with a smooth distribution on degree scales.

Any component of 
the gamma-ray flux which is extended on the scale of a few
tenths of a degree or larger should not show any observable time variation,
due to the large variation in path lengths. Burst-like 
emission behavior of the source is completely averaged out.
Therefore, the presence of an extended halo component should
cause a flux dependence (and time dependence)
of the angular distribution of gamma rays;
for high flux, the point like source component dominates, whereas
for the quiescent flux the extended halo is enhanced.

In a first search for indications of a halo, the ratio
of the event rates in different annular regions around the source was
studied as a function of time. Specifically, the flux in 
various angular ranges
$\theta_1$ to $\theta_2$ 
relative to to direction to Mkn 501 was compared to the flux in the central
peak, $\theta < 0.05^\circ...0.1^\circ$. No statistically significant evidence
of a time dependence of the ratios was observed.

In the following, we will concentrate on the flux dependence of
angular distributions and specifically compare the high-flux burst periods
with the quiescent state of the source.
Fig.~1
shows the angular distributions of gamma-rays from Mkn 501, 
after background subtraction,
both for the 1997 high-flux burst period
(Aharonian et al 1999a, 1999b), and for the 1998/99
quiescent state (Aharonian et al. 2000a), excluding the flare on
June 26-28, 1998 (Sambruna et al. 2000).
We note that with the event selection used here, the detection
efficiency of the telescope system, i.e. its effective area, is 
constant to better than $\pm 3\%$ for $\theta$ between $0^\circ$
and $0.5^\circ$.
\begin{figure}[htb]
\begin{center}
\mbox{
\epsfxsize6.5cm
\epsffile{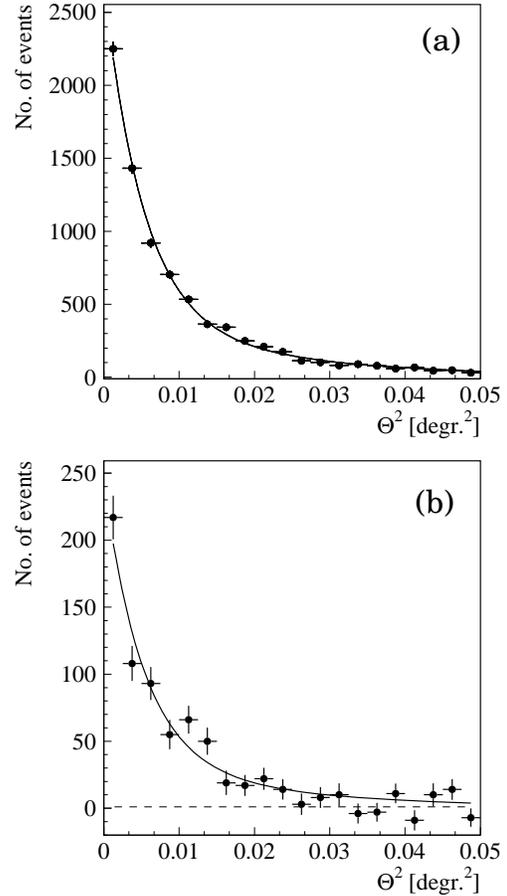}}
\label{fig_halo1}
\caption{Angular distribution dN/d$\theta^2$ of gamma rays 
relative to the direction of Mkn 501. (a) For the 1997 burst
period, and (b) for the 1998/1999 quiescent state, excluding
a short flare in June 1998. Background is subtracted using an
off-source region separated by $1^\circ$ in declination from
the source region. The lines indicate a fit by a superposition
of two Gaussian distributions in $\theta$; the same shape is
used for both data sets.}
\end{center}
\end{figure}
In the angular distributions for the high-state and quiescent-state
angular distributions 
(Fig.~1(a) and (b), resp.), 
no dramatic difference is observed.
The distributions shown in Fig.~1 are not exactly Gaussian,
reflecting the fact that the contributing
events have angular resolutions varying between $0.04^\circ$ and 
$0.10^\circ$. The distributions are well described by a sum of two Gaussian
distributions in $\theta$ (solid lines).

To characterize the angular distribution as a function
of flux, data were binned according to the average flux
measured in each night of observations, and the distribution
in $\theta^2$ was determined for each bin independently.
Six flux bins were used for the 1997 burst data; a seventh
``bin'' contains the 1998 quiescent-phase data, excluding the
June 26-8 flare. The average flux in 1998 is below the flux
of the lowest 1997 flux bin.
Two quantities characterizing the distribution
were examined for each flux bin: the width of the central
peak of the distribution, approximated by a single Gaussian 
(Fig.~2(a)) 
and the yield of gamma-ray events detected in the angular
range between $0.22^\circ$ and $0.50^\circ$ from the direction to the source,
normalized to the yield inside $0.22^\circ$
(Fig.~2(b)).
The first quantity is sensitive to a halo with an angular width
on the same scale as the angular resolution of the instrument,
the second to a halo component with a width significantly 
larger than the angular resolution. Neither quantity shows a
significant dependence on the flux, and the measured values are consistent
with results of simulations of the point spread function of the HEGRA
instrument, using the measured energy spectrum as an input.  
\begin{figure}[htb]
\begin{center}
\mbox{
\epsfxsize6.5cm
\epsffile{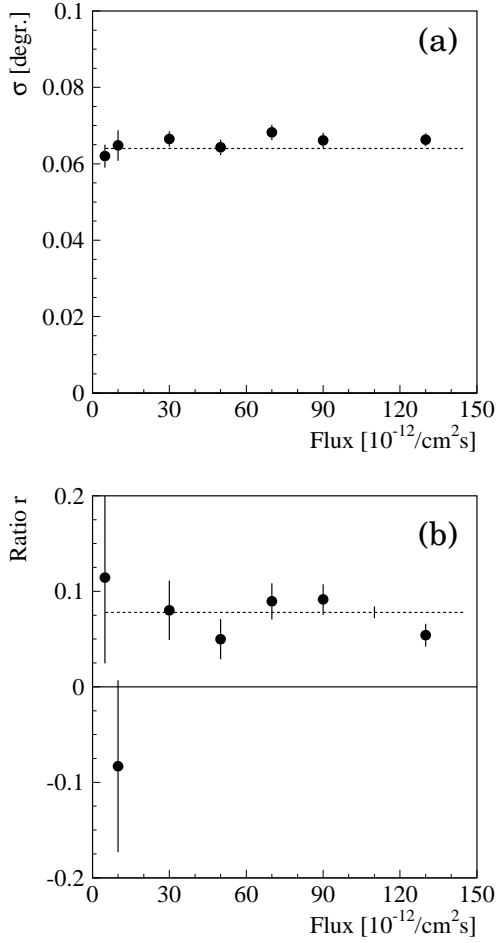}}
\label{fig_halo12}
\caption{(a) Gaussian width of the angular distribution of
gamma-rays relative to the direction to Mkn 501, as a function
of the gamma-ray flux. The point for the lowest flux is obtained from
the 1998/99 quiescent-state data; the points at higher flux
are based on data
from the 1997 outburst. The dashed lines indicates the result of
the simulation.
(b) Flux in the angular range $0.22^\circ$ to 
$0.5^\circ$ relative to the source, normalized to the 
flux inside $0.22^\circ$. Background is subtracted using an
off-source region separated by $1^\circ$ in declination from
the source region. The error bar on the dashed line indicates
the statistical error of the simulation results.}
\end{center}
\end{figure}

In order to provide a flux limit on the halo flux, a model
for the shape of the halo is required. Two variants were
investigated. 
\begin{figure}[htb]
\begin{center}
\mbox{
\epsfxsize6.5cm
\epsffile{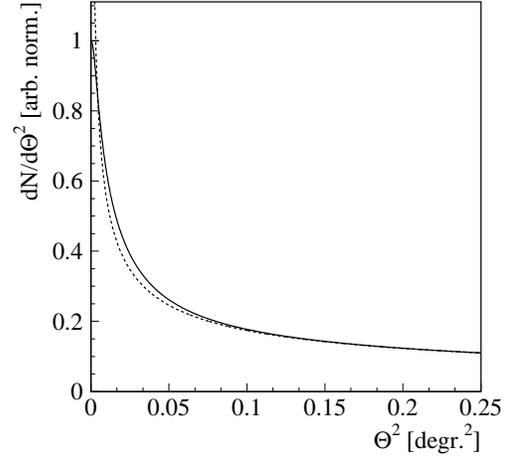}}
\label{fig_halo5}
\caption{{\bf Expected}
angular distribution of halo photons for the assumption
$N(\theta) \sim \theta$, equivalent to 
$\mbox{d}N/\mbox{d}\theta^2 \sim 1/\theta$. The dashed line shows
the original distribution, the full line the distribution
resulting once the angular resolution of the instrument is folded in.}
\end{center}
\end{figure}
\begin{itemize}
\item A Gaussian distribution, 
$\mbox{d}N/\mbox{d}\theta^2 \sim \exp(-\theta^2/2 \sigma_{Halo}^2)$.
\item The distribution 
$\mbox{d}N/\mbox{d}\theta^2 \sim 1/\theta$ 
(Fig.~3).
The halo simulations of Aharonian et al. (1994, 1995) 
typically show, for
small angles, a distribution $N(\theta) \sim \theta$ where 
$N(\theta)$ is the number
of halo photons inside a cone of opening angle $\theta$,
equivalent to 
$\mbox{d}N/\mbox{d}\theta^2 \sim 1/\theta$.
\end{itemize}
To derive limits, the angular distribution of the quiescent-state
data
was fit by a linear combination of the high-state angular
distribution, and a halo contribution. 
The fit covers the range in $\theta^2$ from 0 to 0.25.
For none of the halo
models, a significant halo contribution was found, and we
give in 
Fig.~4 
the 99\% upper limits on the halo flux within
an (half-)opening angle of $0.5^\circ$ around the source
(solid line). The
fit and the quoted limits take into account the spill-over 
of signal photons into the background region for wide halos.
The flux limits for the Gaussian halo are lowest for the
region of intermediate halo scales, $\sigma_{Halo} \approx 0.1^\circ$, 
and deteriorate
both for narrow halos, which cannot be distinguished from
the direct flux, and for very wide halos, where the spill-over
into the background region limits the sensitivity.
The dashed line in 
Fig.~4
 indicates the limit
obtained for the halo model with $N(\theta) \sim \theta$, see 
Fig.~3. A halo flux corresponding to the limit would roughly
triple the expected number of events in the angular range
from $0.22^\circ$ to $0.5^\circ$, see Fig. 2(b).
All limits assume that the energy spectrum of cascade photons
resembles the spectrum measured for the quiescent state, i.e.
show a power law with an index around 2 and a cutoff at a few TeV.
\begin{figure}[htb]
\begin{center}
\mbox{
\epsfxsize6.5cm
\epsffile{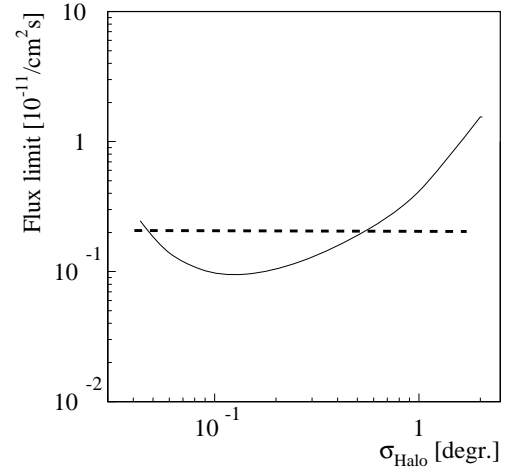}}
\label{fig_halo3}
\caption{Full line: upper limit (99\% C.L.) on the halo flux
above 1 TeV, 
integrated up to $0.5^\circ$ from the source, assuming
a Gaussian distribution of halo photons,
$\mbox{d}N/\mbox{d}\theta^2 \sim \exp(-\theta^2/2 \sigma_{Halo}^2)$
as a function of the width $\sigma_{Halo}$.
Dashed line: limit for a distribution 
$\mbox{d}N/\mbox{d}\theta^2 \sim 1/\theta$.}
\end{center}
\end{figure}

As discussed earlier, a more model-independent limit on the halo flux can be
obtained by using a distant region for background subtraction, rather
than the region displaced by only $1^\circ$ in declination. Data taken
on other objects can, e.g., serve as a background region. The difficulty
is that already minor changes in the 
performance or calibration of the instrument can change the background
yield. As an example, 
Fig.~5(a) 
shows the distribution
in the {\em mean scaled width} of images for a halo region in the
Mkn 501 quiescent-state data, and in an equivalent region in Crab data
taken at a different time. In this case, the `halo region' was defined
as a ring around the source, with an inner radius of $0.33^\circ$
and an outer radius of $1^\circ$. Together with the required angular
resolution of better than $0.1^\circ$, the inner cut ensures that virtually
no direct gamma-rays spill over into the background region. 
Fig.~5(b) 
illustrates, for comparison, a weak gamma-ray
signal peaking at a {\em mean scaled width} of 1.
While the two distributions shown in 
Fig.~5(a) 
are quite
similar, there are differences in the shape, both in the gamma-ray
region ({\em mean scaled width} between 0.8 and 1.2) and outside. 
The differences can be explained by systematic shifts of the
distribution by as little as 0.01 to 0.02 units, caused e.g.
by changes in the alignment of the telescope mirrors with time.
The maximum allowable number of gamma-rays in 
the halo region depends on exactly how the two distributions are normalized
to each other, before performing the background subtraction. 
Worst-case assumptions allow a photon flux of $1.1 \cdot 10^{-11}$/cm$^2$s
above 1 TeV
in the angular range between $0.33^\circ$ and $1.0^\circ$ from Mkn 501,
equivalent to $1.3 \cdot 10^{-8}$/cm$^2$s~sr. Work is in progress to better
understand the differences in the distributions, to allow a more reliable
background subtraction.
\begin{figure}[htb]
\begin{center}
\mbox{
\epsfxsize6.5cm
\epsffile{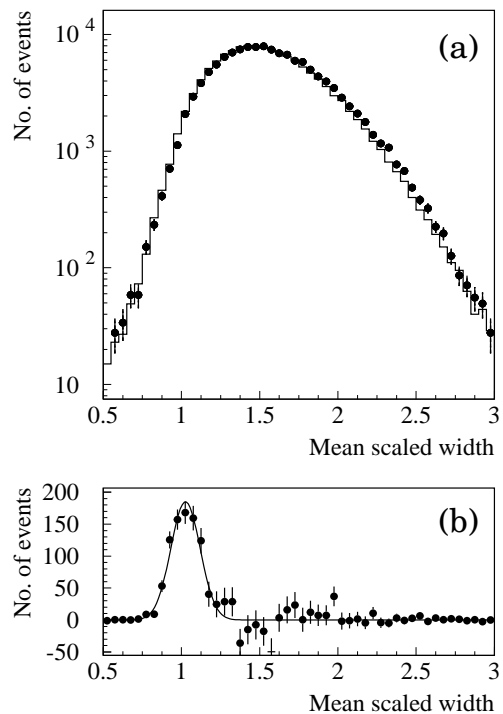}}
\label{fig_plota}
\caption{(a) Distribution in the {\em mean scaled width} of images 
in the halo region of the quiescent-state Mkn 501 data sample (full line),
compared to the same distribution for the Crab data sample,
scaled up to the Mkn 501 data. The halo
region covers angles between $0.33^\circ$ and $1^\circ$ relative to the
source. (b) Distribution for gamma rays, in this case for the
quiescent-state Mkn 501 data, after background subtraction.}
\end{center}
\end{figure}

\section{Concluding remarks}

The search for a pair halo surrounding the direct TeV
gamma rays from Mkn 501 shows no indications for a halo;
the angular distribution in the quiescent phase -- where
a halo should be best visible -- is within errors identical
to the distribution during the burst phases. The upper
limits on a halo contribution depend on the model for 
the angular distribution of halo gamma rays; both very
narrow and very wide halos are difficult to detect. In
the region of optimal sensitivity, the upper limit on
the halo flux within $0.5^\circ$ from the source is around
1/1000 of the peak flux during bursts. For conventional
halo models (Aharonian et al. 1995, 1994), and `nearby'
sources such as Mkn 501, the flux within $0.5^\circ$
is $O(10\%)$ of the total halo flux (for the more realistic `high IR'
variant of the model),
so that the total halo flux is limited to $O(0.01)$ of the peak
burst flux. This value is above the expectations on the
basis of energetics, and does not seriously constrain 
models unless a very large duty cycle or a very poor beaming
of the direct radiation is assumed. The situation changes
if one assumes the dramatic absorption of gamma rays above
10 TeV, as implied by recent measurements of the density of the
infrared background radiation; such scenarios imply a huge source
power at very high energies and correspondingly a strong halo.

A narrow halo with a width in the $0.1^\circ$ range would be
generated if intergalactic magnetic fields are very small,
in the range below $10^{-16}$ G (see also Plaga 1995). 
Detailed cascade simulations will be required
to determine which range of model parameters can be excluded; 
this goes beyond the scope of
the present paper.

\section*{Acknowledgments}

The support of the HEGRA experiment by the German Ministry for Research 
and Technology BMBF and by the Spanish Research Council
CYCIT is acknowledged. We are grateful to the Instituto
de Astrof\'\i sica de Canarias for the use of the site and
for providing excellent working conditions. We gratefully
acknowledge the technical support staff of Heidelberg,
Kiel, Munich, and Yerevan. GPR acknowledges
receipt of a Humboldt Foundation postdoctoral fellowship.

\end{document}